\documentclass[runningheads]{llncs}
\usepackage{graphicx}
\usepackage{epsfig,endnotes,xspace}
\usepackage{amsmath}
\usepackage{versions}

\usepackage[version=0.96]{pgf}
\usepackage{tikz}
\usetikzlibrary{arrows,shapes,backgrounds,positioning}

\newtheorem{thm}{Theorem}

\newcommand{\sysname}{Blockmania\xspace}

\excludeversion{longdoc}

\includeversion{names}

\begin{document}

%

\title{\sysname: from Block DAGs to Consensus }
\begin{names}
\subtitle{DRAFT (v0.5, 25 Sept 2018)}
\end{names}
%
%
\author{}
\institute{}

\begin{names}
\author{George Danezis\inst{1,2} \and
David Hrycyszyn\inst{2,3}}
\authorrunning{G. Danezis et al.}
%
\institute{
University College London (UK)
\and
{\tt chainspace.io}
\and
{\tt vegaprotocol.io}
}
\end{names}

\maketitle              
\begin{abstract}
\sysname is a byzantine consensus protocol. Nodes emit blocks forming a directed acyclic graph (block DAG) that is subsequently interpreted by each node separately to ensure consensus with safety, liveness and finality. The resulting system has communication complexity $\mathcal{O}(N^2)$ even in the worse case, and very low constant factors --- as compared to $\mathcal{O}(N^4)$ for PBFT; it is leaderless; and network operations do not depend on the composition of the quorum or node stake. This makes \sysname very efficient (leading to over 400K transactions per second on a wide area network), and ideal for dynamic membership and flexible and non-interrupted proof-of-stake protocols. A X-\sysname variant, has $\mathcal{O}(N)$ communication cost but also higher latency $\mathcal{O}(\log N)$. 
\begin{longdoc}
\end{longdoc}
\end{abstract}

\section{Introduction}

Distributed ledgers, such as IOTA\footnote{\url{http://iotatoken.io/}} and Byteball\footnote{\url{https://byteball.org/}}, rely on nodes creating a directed acyclic graph of blocks (Block DAG): nodes accept transactions and other blocks, and periodically seal them into a sequence of their own blocks. Mechanisms, such as The Tangle~\cite{popov2016tangle}, are then employed to achieve probabilistic consensus and finality. However, can such a Block DAG provide the basis for consensus with more traditional safety, liveness and finality properties?

We propose a new paradigm for designing secure distributed algorithms. Nodes jointly create a Block DAG, and subsequently interpret it as a byzantine fault tolerant consensus mechanism. Thus two honest parties with a partial view of the Block DAG, can reach agreement on which blocks will be accepted by all. As for traditional PBFT~\cite{castro1999practical} safety is guaranteed under asynchrony, and liveness under partial synchrony~\cite{dwork1988consensus}.

Specifically, our contributions include: (1) an efficient block gossip protocol, and its lightweight validity rules;
(2) a simplified leaderless byzantine consensus protocol;
(3) how to interpret the gossip protocol as an instance of the consensus protocol, and the security and efficiency advantages of this approach;
(4) economic tools to secure the system, including an ordering based on fees and a delegated proof-of-stake quorum system;
(5) a theoretical analysis of its costs, and concrete performance measurements on local and wide area networks; 
(6) X-\sysname that has lower, $\mathcal{O}(N)$, communication cost. Proof for all theorems are in the appendix.

\section{Network operation through Block Gossip}

We first describe the network layer of \sysname. Nodes are network servers composing the infrastructure of \sysname. We assume that nodes possess a signature key, using an asymmetric signature scheme, and all others can authentically verify their messages using the corresponding verification keys. Clients are external to nodes and emit transactions that need to be agreed upon. The goal of the \sysname protocol is to ultimately ensure that all honest nodes arrive at the same ordering of transactions (safety) despite a subset of nodes being byzantine; and for the protocol to finalize an ordering of transactions from honest nodes, despite the actions of byzantine nodes (liveness).

\vspace{3mm} 
\noindent {\bf Core networking.} The basic inter-node network operation in \sysname consists of each node periodically broadcasting a block to all other nodes. The nodes are loosely synchronized, using for example a byzantine clock synchronization protocol~\cite{lamport1984byzantine}, and emit blocks at regular intervals. Figure~\ref{fig:block} illustrates the structure of those blocks. 
\begin{figure}[t]
\begin{minipage}[t]{0.5\textwidth}
\begin{align*}
    &\mathbf{datatype}\ block = \{\\
    &\quad n:\ \textit{node identifier}; \\
    &\quad k:\ \textit{integer};\\
    &\quad \textit{prev}:\ \textit{hash\ of\ block};\\ 
    &\quad \textit{entries}:\ \textit{list\ of\ element};\\ 
    &\quad \textit{sig}:\ \textit{signature};\quad \}
\end{align*}
\end{minipage}
\begin{minipage}[t]{0.5\textwidth}
\begin{align*}
    &\mathbf{datatype}\ element = \\
    &\quad |\ \textit{Transaction\ of} \\
    &\qquad \textit{\{ t:\ opaque; fee:\ integer; \}} \\
    &\quad |\ \textit{Reference\ of} \\ 
    &\qquad \textit{\{reference:\ hash\ of\ block; \}}\\
    &\\
    &\textit{hash(b:block)} \rightarrow \textit{hash\ of\ block};
\end{align*}
\end{minipage}%
    \caption{Data structures and functions used by block exchange to form the Block DAG.}
    \label{fig:block}
\end{figure}

Blocks from each node form a sequence and each refers to the hash of the previous one ($prev$), forming a Hash Chain~\cite{merkle1989certified}. A block includes information about its creator node ($n$), its sequence number ($k$) --- and we say this is a candidate block for position ($n$, $k$). It also contains a sequence of \emph{entries} representing the content of the block.  Each entry is either a transaction received from a client, with an associated fee, or a reference to the hash of a valid block received from another node. The full block is signed using the private key of its creator, and broadcast to all other nodes. Besides broadcasting blocks, nodes listen to requests from other nodes for specific blocks, and respond by sending them the full contents of the block requested.

Clients may connect to any node and submit transactions they wish to process. They can subsequently observe the consensus to determine whether their transaction was included and re-transmit it (possibly to a different node) if it is not. An honest node includes all transactions received from clients as entries into a block it emits.

A node includes references to \emph{valid} blocks received from other nodes as entries. References to a block are computed using a secure cryptographic hash function over the content of the block (excluding the signature).  


\vspace{3mm}
\noindent {\bf Block validity conditions.} A block $\textit{B(n,k, prev, entries, sig)}$ received by a node, is valid under the following conditions:
\begin{enumerate}
    \item The block $B$ has been fully received and stored by the node;
    \item The signature (\emph{sig}) verification succeeds for the creator $n$;
    \item The previous (\emph{prev}) block is valid, and its sequence number lower than $k > 0$. Or if $k=0$, the previous block reference is nil.
    \item All blocks referenced in \emph{entries} are valid.
\end{enumerate}
These validity conditions ensure that a node only includes a reference to a block in its own block, if it has also received and checked for validity all blocks it references, leading to the following properties:
\begin{thm}[DAG Availability 1]\label{dag:avail1}
An honest node has received and stored the full block DAG, starting from any genesis blocks, to the last block it emits.
\end{thm}

\begin{thm}[DAG Availability 2]\label{dag:avail2}
If an honest node emits a block, all honest nodes will eventually receive and consider valid this block, and all blocks referenced directly or indirectly by this block.
\end{thm}

The validity conditions do not, by themselves, require or ensure consensus about which block corresponds to a particular position. A dishonest node $n'$ may contradict itself (i.e.\ equivocate), and send two distinct blocks $B_1$ and $B_2$ for a single position ($n'$, $k'$). Similarly, a byzantine node $n'$ may not send a block for a position ($n'$, $k'$), for some time, or ever. Honest nodes reference all valid blocks received in their own blocks, including contradictory ones, and proceed with the protocol.

It is the function of the next phase of \sysname, namely \emph{interpretation}, to ensure \emph{terminating reliable broadcast semantics} for each position ($n$, $k$), as well as a \emph{consensus on full ordering} of transactions within those blocks.

\section{From Block DAGs to Position-level Consensus}

The aim of the interpretation phase is for all honest nodes, despite holding a different subset of the block DAG at any point, to eventually agree on a specific block for any position, namely $(n, k) \rightarrow B_{n,k}$; or alternatively agree that there is no such block, namely $(n, k) \rightarrow \mathit{nil}$. The latter could be due to byzantine nodes propagating conflicting information or no information. The terminating nature of the broadcast makes this protocol equivalent to byzantine agreement. 


The terminating reliable broadcast protocol we develop is a simplified variant of PBFT~\cite{castro1999practical}, where: (1) only a single decision, rather than a sequence, has to be agreed on for any position $(n, k)$;
(2) it is egalitarian and leaderless in its operation; 
(3) we do away with PBFT `checkpointing', that is very complex.
As in PBFT the operation of nodes is deterministic given the messages they receive.

Similar to other works in byzantine fault tolerance, we assume reliable transmission between honest nodes---a byzantine network may delay arbitrarily blocks but eventually they are to be delivered. In practice we achieve this through re-transmissions. Each honest node monitors the chains of other nodes. In case their own blocks are not included into those within an estimated round trip time, they may re-transmit them. Further, an honest node may request missing blocks from other nodes, when those have been included into their chain as valid (see Theorem~\ref{dag:avail2}).

\subsection{Abstract terminating reliable broadcast protocol} \label{atrb}

We first present the abstract terminating reliable broadcast protocol. This protocol is abstract, in the sense that \emph{it is never materialized as real-world messages} exchanged by nodes. Instead, the structure of the block DAG itself is interpreted by each node, to infer those protocol messages and drive a state machine, as discussed in section~\ref{interpret}.


The terminating reliable broadcast protocol relies on a quorum of nodes exchanging messages to achieve consensus on a value. The set of nodes must be defined and known to all; a quorum of $3f+1$ nodes can tolerate up to $f$ byzantine nodes. We stress that the gossip protocol by itself does not require nodes to specify the composition of such a quorum. For the moment we will assume the set of nodes is externally specified, but in Section~\ref{stake} we discuss how to compose such quorums through a proof-of-stake scheme.

As for PBFT the abstract messages exchanged are \emph{pre-prepare}, \emph{prepare}, \emph{commit}, \emph{view-change} and \emph{new-view} (there are also no messages relating to checkpointing). Each node maintains some state per decision it needs to reach: namely, a current view number $\mathit{view}_n$, a list of input messages $\mathit{in}_n$ and output messages $\mathit{out}_n$. The initial view number is zero and the message sets are empty.

The abstract protocol for deciding on position $(n, k)$, assuming no timeouts, proceeds as follows:
\begin{itemize}
    \item A block originator node $n$ broadcasts a $\textit{pre-prepare}(n, k, B, v=0)$ to all nodes proposing block $B$ as a candidate for position $(n,k)$. The value $v$ is the view sequence number, and it starts with zero. Honest nodes only propose one block for any given position, and never equivocate.
    
    \item Upon receiving the first $\textit{pre-prepare}$ for a view $v$, a node with $\mathit{view}_n$ equal to $v$, broadcasts $\textit{prepare}(k, n, B, v)$. If $v=0$ the signature of $n$ is checked before accepting the message. A node will not \emph{prepare} for a different block in that view, and includes the $\textit{pre-prepare}$ and $\textit{prepare}$ message into its $\mathit{in}_n$ set. 
    
    \item Nodes await $\textit{prepare}(k, n, B, v)$ messages, and add them to $\mathit{in}_n$ if they are received within $\mathit{view}_n = v$. Upon receiving $2t$ $\textit{prepare}(k, n, B, v)$ messages, and a corresponding \emph{pre-prepare} message  in $\mathit{in}_n$, a node broadcasts $\textit{commit}(k, n, B, v)$.
    
    \item Nodes await $\textit{commit}(k, n, B, v)$ messages and include them in $\mathit{in}_n$, if $\mathit{view}_n \geq v$. Upon receiving $2t+1$ $\textit{commit}(k, n, B, v)$ messages, the node considers the value $B$ (which may be `nil') to be the value \emph{decided} for position $(n, k) \rightarrow B$, but continues to participate in the protocol.
\end{itemize}
However, due to network delays, or malicious nodes' actions, a timeout may occur in any view $\mathit{view}_n$. Timers are set when the view is entered, and expire if a block for position $(n, k)$ is not agreed within a certain logical time window within the view. In such cases, a view-change is triggered, which moves the state machine to a new view:

\begin{itemize}
    \item At any time, within a view $\mathit{view}_n$, a node may emit $\textit{view-change}$, and increase its view to $\mathit{view}_n + 1$. If the node has sent a commit message for $B$, it sends $\textit{view-change}(k, n, \mathit{view}_n, P)$ where $P$ is the set of all prepare and pre-prepare messages supporting value $B$ as the block for $(n, k)$ (for the highest previous view a message was committed). Otherwise, if no value has ever been committed, the set is left empty, as $P = \{\}$. The node stops accepting any new messages for past views except for $\textit{commit}$; and sends no messages for past views.
    
    \item Nodes await $\textit{view-change}(k, n, v, P)$ messages, and include them in $\mathit{in}_n$, if they are correctly formed and $v > \mathit{view}_n$. Upon receiving $2t+1$ view-change messages with view $v > \mathit{view}_n$, a node sets $\mathit{view}_n = v$. It broadcasts $\textit{new-view}(n, k, \mathit{view}_n, V)$. The set $V$ contains all $2t+1$ view-change messages on which the new-view message is based.
    
    \item Upon receiving the first new-view message $\textit{new-view}(n, k, v, V)$, with $v \geq \mathit{view_n}$ and $v > 0$, it sets $\mathit{view}_n = v$. The node establishes whether any view change messages in $V$ commits to a block $B$. If so it interprets the message as being a \emph{pre-prepare} for this block in the new view, namely $\textit{pre-prepare}(n, k, B, v)$. Otherwise, it interprets it as $\textit{pre-prepare}(n, k, nil, v)$ and responds accordingly.
    
    \item The protocol proceeds through $\textit{prepare}$, $\textit{commit}$ phases or subsequent view changes as above.
\end{itemize}
Any message broadcast or sent by a state machine is also included in the $\textit{out}_n$ buffer. The protocol guarantees safety and liveness properties within a byzantine threat model, and assuming partial synchrony:
\begin{thm}[Terminating Reliable broadcast (safety)]\label{trb:safe}
If two honest nodes reach a decision for a position ($n$, $k$), it will be the same decision, assuming at most $f$ byzantine nodes.
\end{thm}
\begin{thm}[Terminating Reliable broadcast (liveness)]\label{trb:live}
A decision will eventually be reached for any position ($n$,$k$) by all honest nodes, assuming at most $f$ byzantine nodes and partial synchrony. If $n$ is honest, it will be for the block $n$ prepared for position ($n$, $k$).
\end{thm}

\subsection{Block DAGs as Terminating Reliable Broadcast.}
\label{interpret}

The previous section describes the \emph{terminating reliable broadcast} protocol as a state machine consuming and emitting messages to determine a consensus decision for position $(n,k)$. However, the messages that drive the state machine in \sysname are not materialized as actual network messages. They are instead inferred independently by each node, from the their local copy of the block DAG. 


A node $n'$ interprets each valid block $B_{(n, k)}$ as \emph{carrying a set of state machines} for non decided positions. The state machines are copied from the previous block in the chain from node $n$ (variable \emph{prev} in the block), or are an empty set for $k = 0$. A state machine for decision $(n', k')$ is initialized within a block $B_{(n, k)}$, under three conditions: (1) for the block itself if block $(n', k') = (n, k)$; (2) when the first block for $(n', k')$ is included in the block $B_{(n, k)}$; or (3) within the block $B_{(n', k')}$ when $k' = k$ (upon which just a timeout is defined). Each block is also associated with a set of outgoing messages, that correspond to the union of messages in the $\textit{out}$ sets of all state machines contained within the block.

We interpret a block $B_{(n, k)}$ including a reference to a block $B_{(n',k')}$, as node $n'$ sending to the receiver $n$ the messages included in the $\textit{out}$ buffer of included block $B_{(n',k')}$. Those messages drive the state machines within block $B_{(n, k)}$. The sequence of blocks included within $B_{(n, k)}$, denotes the exact order in which messages are received from other blocks. As a result the state for all decisions within block $B_{(n,k)}$ evolves, and new messages may be added to the block $\textit{out}$ buffer.

To prime a state machine for decision $(n',k')$ we inject within the state machine of any block $B_{(n', k')}$ a $\textit{pre-prepare}(n',k', B_{(n',k')}, v=0)$ message, and also include it in its $\textit{out}$ buffer. Namely, a block claiming to be for position $(n', k')$ initializes the terminating reliable broadcast with a $\textit{pre-prepare}$ message for itself. Once a state machine within the DAG of valid blocks reaches a decision to deliver a block $B_{(n',k')}$ for position $(n',k')$ then it is considered decided by the node performing the interpretation.

\begin{figure}[t]
\centering
\begin{tikzpicture}[node distance=1.3cm,>=stealth',bend angle=45,thick,scale=0.45, every node/.style={scale=0.7},auto]

\tikzstyle{transition}=[rectangle,thick,draw=black!75,
  			  fill=black!20,minimum size=9mm, text width=1cm]

\begin{scope}
\input{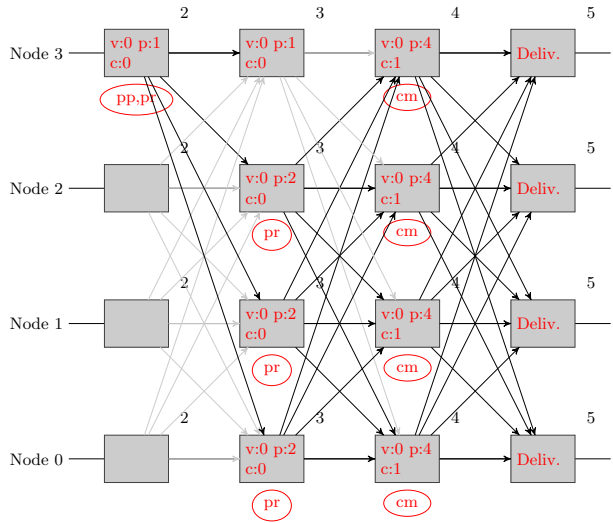}
\end{scope}

\end{tikzpicture}
  \caption{An illustration of a block DAG (black and gray), and its interpretation as a state machine per block for position $(n=3, k=2)$. Each state machine for $(n=3, k=2)$ includes a view number ($v$), and a count for $\textit{prepare}$ ($p$) and $\textit{commit}$ ($c$) messages received by the block, and is illustrated within each block in red (along with a $\textit{Deliv.}$ delivery decision being reached). The $\textit{out}$ buffer for each block for this decision, is summarized in a red circle below the corresponding block (`pp' for $\textit{pre-prepare}$), `pr' for $\textit{prepare}$ and `cm' for $\textit{commit}$. None of the interpreted information (in red) is materialized as network communication; only the blocks are broadcast.}   \label{fig1}
\end{figure}

\vspace{3mm}
\noindent {\bf Illustrated example.} Figure~\ref{fig1} illustrates an example of interpreting a block DAG to reach a decision for round $k=2$ from node $n=3$. All nodes are honest and live (there is no equivocation or missing blocks). We only illustrate the state machine for decision $(n=3, k=2)$ even though state machines for other decisions are also included in each block.

The state machine for $(3,2)$ is initialized through a $\textit{pre-prepare}$ message when the block $B_{(3,2)}$ is created. This results in a $\textit{prepare}$ message within block $B_{(3,2)}$. This block is then forwarded to nodes 1, 2 and 0 in order, leading their state machines to initialize for position $(3,2)$ and also be augmented with an additional $\textit{prepare}$, that are included in their out buffers. Nodes 0 and 3 are first to receive $2f+1 = 3$ $\textit{prepare}$ messages and emit $\textit{commit}$ messages in their $\textit{out}$ buffers. Those are propagated to blocks from all other nodes, that reach a ``deliver'' decision.


\vspace{3mm}
\noindent{\bf Security properties of the interpretation approach.} Interpreting a block DAG, instead of nodes exchanging messages pertaining to the consensus explicitly, ensures some key security properties. 

Transmitting the block DAG, without explicit messages, restricts byzantine nodes' ability to influence honest nodes state machines. The only actions byzantine nodes can perform are: (1) not transmitting blocks for some of their position, (2) delaying arbitrarily the timing of the transmission of blocks for a specific position of theirs, (3) transmitting more than one block for one of their positions, and (4) forming their blocks arbitrarily (irrespective of validity rules). Digital signatures prevent byzantine nodes from forging the transmission of blocks for positions of honest nodes. Validity rules prevent them from having blocks included into honest node block chains, unless those honest nodes have seen the full DAG leading to these blocks, and have checked them as valid.

An honest node interprets the block DAG for valid blocks included in its own blocks. Since the node is honest the interpretation steps are performed correctly, the messages inferred from each state machine are correct, and all state machines for its own blocks and other nodes' blocks are updated correctly. Therefore, there is no need for any logic checks for the node performing the interpretation they are correct: such as the evidence necessary within $\textit{view-change}$ and $\textit{new-view}$ messages. Since those are generated and used by the same honest node they are guaranteed to be correct. This greatly simplifies the implementation of the state machine, and its communication efficiency.\footnote{The casual reader may not appreciate this simplification, unless they have attempted to implement PBFT. The predicates~\cite{castro1999correctness}(page 8) to decide whether a view-change is valid are composed on 5 lines of dense first order logic (see \emph{correct-view-change} and \emph{last-prepared}), while the predicates for checking new-view messages span 6 lines (see \emph{correct-new-view}). For comparison, all other predicates are just 12 lines in total.}
\begin{thm}[Safe Interpretation 1] \label{int:safe1}
Two honest nodes interpreting a block they both consider valid, using the same quorum of nodes, will interpret it in an identical manner. Namely, they will derive the same state for all state machines associated with the block, and the same set of messages. 
\end{thm}
Once a node interprets any state machine as delivering a block for a position, it can stop any further interpretation and consider that all honest nodes will decide the same block for this position. 
\begin{thm}[Safe Interpretation 2] \label{int:safe2}
Assuming at most $f$ byzantine nodes and a quorum of $3f+1$ nodes. Once an honest node interprets a state machine within any valid block as having reached a decision for a position, any other DAG of valid blocks interpreted by an honest node, with the same quorum, will eventually reach the same decision.
\end{thm}

The proof of this theorem (see appendix) makes a very pessimistic assumption, that byzantine nodes may force honest nodes to interpret messages from their blocks as arbitrary terminating reliable broadcast protocol messages. In reality, they may only force honest nodes to interpret such messages as a messages resulting from a valid run of a state machine --- potentially driven by an arbitrarily chosen set of input blocks. This may allow further optimizations, that we do not explore in this paper.

Split view, or equivocation attempts, in which dishonest nodes advertise more than one block for a single position, are also guaranteed to be eventually detected if they ever have an impact on honest nodes. Consider two honest nodes including into their chains two different blocks pertaining to the same position: say $B_{(n,k)}$ and $B'_{(n,k)}$. Subsequently, and as soon as the two honest nodes exchange a block, they will detect the contradiction, and have conclusive evidence node $n$ is byzantine. Honest nodes may stop including blocks from such byzantine nodes into their DAG after detecting such an event: as long as there are $2f+1$ honest nodes, this does not impact safety or liveness. Eventually, all honest nodes will exclude such a byzantine node.

\vspace{3mm}
\noindent{\bf Timers and timeouts.} We use logical timers to ensure liveness. As soon as a state machine at block $B_{(n,k)}$ is initialized for deciding position $(n',k')$ with $\textit{view}_n = 0$ a timeout is set, for some subsequent round $k+T$. These timeouts are part of the state of block $B_{(n,k)}$. The timeout is triggered when block $B_{(n, k+T)}$ is created by $n$. If no decision has been reached before this block for $(n',k')$, then node $n$ includes a $\textit{view-change}$ message for this state machine, and initializes a new timeout.

The value of $T$ is determined dynamically to ensure conditions of partial synchrony are eventually reached, no matter the exact maximal network delay bound under eventual synchrony. Each node determines the delay, in blocks, within which it receives blocks from other nodes --- through looking at the round $k'$ of blocks from others it includes in its own block for round $k$ (the delay is $|k' - k|$). It then selects the a value $t$ such that it larger than the lowest $f$ values, and lower than the $f$ largest values -- ensuring it is bound from above and below by the delay from an honest node~\cite{dolev1986reaching}. The value $T$ is then determined by a constant $c > 3$ as $T = c\cdot t$ (we use $c = 10$ in our implementation). This computation depends on the DAG, and is therefore deterministic.

\vspace{3mm}
\noindent \textbf{Separating interpretation from networking.} An important feature of interpreting block propagation as messaging, instead of materializing messages in the network, is that nodes do not have to update the state machine, and send the resulting messages, their normal networking operations. Nodes simply perform the gossip block propagation: they accept transactions from clients and valid blocks from other nodes; reference them into blocks; that they broadcast to all. In particular nodes do not need to know a full list of other nodes participating in the protocol. They also never need to block or wait for $2t+1$ messages to perform any action, as in traditional PBFT and related protocols.

Thus, the protocol above illustrates that \emph{consensus in in the eye of the beholder}: it is only once an observer of the gossip block propagation --- which could be a node or a totally separate entity --- decides the quorum composition and size, that they may interpret the block DAG as terminating reliable broadcast decisions.  Two honest observers of the same DAG will eventually reach the same decision for all positions, as long as they use the same quorum of nodes for their interpretation.

\begin{longdoc}
Interestingly, if two observers chose to interpret the same block DAG using two different quorums of cardinality $3f+1$, say $S_0$ and $S_1$, the terminating reliable broadcast reaches the same decisions if $|S_0 \cap S_1| \geq 2f+1$ and if those overlapping nodes are honest. This follows from the safety properties of the terminating reliable broadcast, that can tolerate up to $f$ byzantine nodes. This is not a property we rely on, but might be of independent interest.
\end{longdoc}

\subsection{Full Consensus With Fee Tie-breakers.}

The terminating reliable broadcast primitive ensures that for a set of participating nodes, and for a sequence number $k$, all nodes will eventually assign a block to each of $(n, k)$ decisions, even if this block is `nil'. We extend this to a consensus protocol to sequence all actions, based on the round of the block that contain them, and using the transaction fee to handle ties.

Each transaction is included in a delivered block $a \in B_{(n, k)}$, and is associated with the round of the block (namely $\rho(a) = k$). Each transaction is also associated with a fee included in the transaction, namely $\phi(a)$. Upon reaching a decision for all $N$ blocks for a round $k$ (one block for each node), nodes order all transactions in all those blocks according to their corresponding tuples $(\rho(a), -\phi(a))$. This ensures that transactions in earlier rounds take priority, and in case of ties transactions with the largest fee take priority.

Once transactions from all blocks in a round $k$ are sequenced in the same way across all honest nodes, they can be executed in order to achieve byzantine fault tolerant replication of any deterministic program, following the well established state machine replication paradigm~\cite{schneider1990implementing}.

\vspace{3mm}
\noindent {\bf Accounts.} All nodes maintain a replicated ledger mapping accounts, referenced by public signature key, and available funds. The fee is a single signed statement that some value $\phi(a)$ should be deduced from some account to support transaction $a$ upon execution. At the stage of ordering, it is not necessary to establish if the funds are available in the account, or even whether the account exists or is valid --- this can be checked at the time of the sequenced execution of the transaction, with those that are invalid or not supported by funds resulting in a no-op.

The resulting ordered sequence of transactions is executed, in the determined order by all nodes. Each transaction in turn is checked for a valid fee, the fee is deduced, and if sufficient funds are available it is executed. To maximize flexibility we do not specify the mechanism by which balances may be increased, which could be external to the consensus (by paying using some fiat currency), or internal to the system through some cryptographic tokens and transactions to move value between accounts. 

\vspace{3mm}
\noindent {\bf Alternative tie breakers.} We note that prioritizing transactions according to a fee deviates from usual BFT protocols, with no notion of a token system or fees. Some may even object that such prioritization unfairly advantages clients with access to more funds to prioritize their transactions. However, we note that a better resourced node can also achieve a similar priority in traditional BFT protocols: they simply need to operate clients from network positions that are closer to the leader (in PBFT). This is also an established strategy in high-frequency trading clients, that are provisioned very close to the central processing nodes. Thus, \sysname simply makes this imbalance visible, by allowing nodes to directly influence the priority of their transactions by expending higher fees, rather than doing so indirectly.

For designs that prefer to not embody an accounting system into their consensus mechanism, a more traditional approach may be used to determine the total order of transactions across blocks in the same round. Those transactions can be ordered by the value of their hash. This mechanism is equivalent to a hashcash~\cite{back2002hashcash} proof-of-work scheme: clients that wish their transaction to take priority will spend resources to mint transactions with lower hashes. 

In systems where transactions encode a causal order, it may be used as part of the sequencing process. In this case, transactions within all rounds agreed are ordered causally, and the fee is only used as a tie-breaker in cases of conflicts that the causal order alone cannot resolve.

\subsection{Embedding into a Proof-of-Stake Consensus System.}
\label{stake}

 \sysname may be used with a proof-of-stake system, to dynamically define the quorum, and has properties that make it very suitable for this purpose. 

\vspace{3mm}
\noindent{\bf Locking state and reaching decisions.} Consensus operates in epochs, of $R$ rounds each. Before the beginning of an epoch each node locks some `stake'. Nodes can freely lock any value from their accounts into stake; such stake remains locked for the subsequent epoch and only gets unlocked at the start of the epoch after. Stake may also be delegated: a node may lock some stake and delegate it to another node to act with the authority of such stake for the epoch.

At the beginning of each period all nodes determine the stake weight of all nodes in the system, denoted as $n \rightarrow \psi(n)$ --- this includes the stake locked by node $n$, as well as any directly or indirectly delegated stake locked by other nodes for $n$. We denote the totality of stake as $\Psi = \sum_n \psi(n)$.

All decisions relating to the interpretation of the terminating reliable broadcast (see section~\ref{atrb}) are interpreted on the basis of the total stake during the epoch. Namely, the decision to issue $\textit{commit}$, is done on the basis of receiving $\textit{prepare}$ messages from nodes representing $\lfloor \frac{2}{3}\Psi + 1 \rfloor$ of all stake; similarly decisions to reach a decision based on \textit{commit} messages, or enter a new view based on \textit{view-change} messages representing that fraction of stake.

Nodes with stake make decisions through \sysname for an epoch, and seal transactions. Upon deciding blocks for all positions up to round $R-1$ of an epoch, they process all transactions relating to locking, delegating and unlocking stake, and update the accounts of stake of nodes for the next epoch. Fees collected over the epoch are re-distributed to the nodes that facilitated the consensus.

\vspace{3mm}
\noindent{\bf Incentives and slashing.} Fees, for executed transactions, are redistributed to nodes according to the blocks that have been decided upon by \sysname. Fees are distributed to the creators of all decided blocks, proportionally to the stake they represent, namely $\psi(n)$ for each block decided from $n$. (We leave it up to those nodes to share that stake with those that may have delegated it to them.)

For each block that is decided as `nil' a fraction of $1/R$ of the stake of the node that should have emitted it is slashed for the next epoch. That stake is redistributed to the other nodes in the same fashion as fees are. If any node has detected emitting two contradictory blocks, all its stake is totally slashed and redistributed to other nodes in full.

\vspace{3mm}
\noindent{\bf Advantages of \sysname.} The network layer, and block gossip, underlying \sysname operates without requiring the interpretations of blocks, the quorum or nodes' stake. As a result, there is no need to interrupt service when the epoch changes, to update those. Nodes can continue the block gossip with each other, and subsequently interpret blocks under the light of the stake of the new epoch, to reach consensus. 

More provocatively, nodes may participate in multiple parallel quorums and proof-of-stake systems simultaneously: each with their own stake tokens and rules. Since network operations do not depend on the interpretation of blocks, nodes may operate continuously, and rely on third parties to interpret the block DAG, execute sequenced transactions, and keep accounting of stake. This represents a \emph{radical separation between the network functions of a node, and the consensus functions} that may be entirely outsourced to different parties.

\section{Performance}

\noindent{\bf How to measure communication cost.} The FLP theorem~\cite{fischer1985impossibility} establishes that consensus is impossible in a fully asynchronous setting, through a deterministic algorithm. \sysname, like PBFT, ensures safety under asynchrony but requires partial synchrony~\cite{dwork1988consensus} for liveness. Therefore, communication costs are measured in the number of bytes that need to be exchanged per consensus decision reached, within a period of synchrony, since under asynchrony a byzantine network can delay decisions indefinitely. We also consider the amortized cost over multiple decisions.

\vspace{3mm}
\noindent{\bf Communication costs.} \sysname's communication complexity is $\mathcal{O}(N^2)$, where $N$ is the size of the quorum. Each block from a node is broadcast to all other $\mathcal{O}(N)$ nodes. Further, all nodes include, at some point, a hash of all other blocks into their own chain, which results in blocks of average size $\mathcal{O}(N)$ --- resulting in an overall $\mathcal{O}(N^2)$ communication cost.

Concretely, the communication cost for each block sealed includes the block meta-data (of fixed size), including a hash to the previous block ($20$ bytes for 80 bit security), and additionally on average a hash to all other blocks ($(N-1)\cdot 20$ bytes) as well as the size of all sealed transactions in the block. The overhead, for 31 nodes ($f=10$) is $20 \cdot 31^2 \approx 19$ kbytes per block; for 1000 nodes ($f=333$) the overhead is $20 \cdot 1000^2 \approx 19$ Mbytes.

It is worth reflecting on how we save $\mathcal{O}(N^2)$ communication cost compared with the PBFT protocol (which has cost $\mathcal{O}(N^4)$ under view change), without sacrificing any of its properties in terms of liveness or safety, including under view change. In traditional PBFT, nodes execute the protocol and directly exchange messages pertaining to commit, view-change and new-view. Those messages need to carry sufficient evidence to convince other nodes they are well formed: for commit and view-change this evidence is $\mathcal{O}(N)$ and for new-view it is of size $\mathcal{O}(N^2)$. In \sysname such evidence does not need to be explicitly sent: each honest node interprets correctly the block DAG, which represents the pattern of message delivery to other nodes, and recreates the correct messages to drive the state of all other nodes in the network --- it does not have to convince itself of the correctness of the interpreted messages --- leading to the reductions in communication cost.

Constant factors in the amortized communication cost of \sysname are also greatly reduced by using useful subsequent blocks, and interpreting them to conclude the consensus on previous blocks. Block exchanges for subsequent blocks are interpreted as prepare and commit, as well as view change and new view phases of the terminating reliable broadcast protocol --- as well as phases of the protocol for subsequent blocks. This leads to the concrete overhead of only 20 bytes per node (for 80 bit cryptographic strength) for each block --- which is minimal. Given this small overhead communication complexity is $20\cdot N^2 + \ell \cdot N + c$, where $\ell$ is the size of transactions in a block and framing, and $c$ a small constant factor. Therefore for smaller quorum size $N$ the cost is dominated by the volume of transactions in each block.

\vspace{3mm}
\noindent{\bf Concrete Performance Measurements.} We implemented a networked prototype of \sysname in \texttt{golang}, and measured its performance under different conditions. Each node generates multiple transactions, of 100 bytes length, that it includes in its own blocks emitted every second. Their volume is controlled by a bang-bang controller aiming to keep latency of consensus under 4 seconds. Operating 4 nodes on a single machine leads to a performance of over 1M transactions per second (tps).

When operating different nodes\footnote{\texttt{n1-standard} instances (4 vCPUs, 15 GB memory, debian-9, with 100GB storage).} within the same data center (Google cloud, London) the performance of \sysname incurs actual TCP networking overheads. For 7 nodes ($f=2$) \sysname can process 650K tps, for 10 nodes ($f=3$) 700K tps, for 13 nodes ($f=4$) 680K tps, and for 16 nodes ($f=5$) 640K tps. The differences are not statistically significant (and depends on the exact placement of virtual machines). We observe that the performance of \sysname remains stable in terms of transactions per second: this is expected as the bottleneck is the capacity of the smallest node to received all blocks, which remains stable. 

We also perform measurements across a wide area network on the Google Cloud platform. The nodes were distributed across Asia (E), North America (NW, W), Europe (W), Australia (SE), and South America (E). For 10 nodes ($f=3$) we observe a performance of 430K tps, 13 nodes ($f=4$) 440K tps, and 16 nodes ($f=5$) 520K tps (the differences are not statistically significant across runs). This compares with an advertised 50K tps for SmartBFT, and 10K tps for Tendermint. We do not observe a significant drop in performance, as we increase the quorum size for small quorums, which validates that the concrete cost is dominated by the volume of transactions exchanged.

\begin{longdoc}

\vspace{3mm}
\noindent{\bf Implementation costs and complexity.} In \sysname all nodes need to independently interpret all valid blocks received, update the state machines they represent and deduce messages in their $\textit{out}$ buffers. Each state machine for a decision is initialized at the latest during the round corresponding to the position to be decided, and can be freed once a decision has been reached. Thus the memory required is proportional to the number of nodes, and the latency of decisions, per valid block that may contribute to a decision.

Separating the network operations of block gossip and consensus results in radically less complex code. The code size for the interpretation of the DAG is 300 lines of golang code, or 300 lines of Python3. Further, those lines do not involve any concurrency, asynchrony or network input-output, making their testing for correctness and performance significantly easier.
\end{longdoc}

\section{$\mathcal{O}(N)$ Communication Cost with X-\sysname. }

Abraham et al.~\cite{abraham2018hot}, pose the question whether there exists a $\mathcal{O}(N)$ communication complexity variant of PBFT as an open problem in the literature on byzantine consensus. Previous systems achieve this for steady-state, and under an honest leader, but incur higher costs when view changes become necessary. 

We present k-\sysname that achieves this bound in even in a worst case setting. This is achieved at the cost of increasing the latency, even under synchrony and when all are honest. For small quorums, we advocate the basic \sysname and for $X \leq 3f+1$ the two designs are equivalent. Latency is, in our view, more sensitive than bandwidth: more bandwidth may be purchased at a cost. But the speed of light within fiber is a hard limit on latency, and therefore lower round complexity leading to lower latency is priceless.

X-\sysname relies on the following insight: in the basic design, a node includes all valid blocks received into their block. This leads to $\mathcal{O}(N)$ block size on average. Including all other blocks is not necessary to satisfy validity. Validity states that all blocks included, and their references, are seen and valid; not that all valid blocks have to be included in the chain. Therefore in X-\sysname we only include a subset of the valid blocks received into blocks.

\vspace{3mm}
\noindent{\bf X-\sysname.} In X-\sysname all nodes emit a block in each round, as in the basic \sysname design. Each node maintains a set of valid blocks received forming the valid block DAG. `Head blocks' in this DAG are those that are not referenced by any other block. Nodes include in each block they emit $X$ of the available head blocks, where $X$ is a constant independent of $N$. Those $X$ heads are chosen at random from the heads available, proportionally to the number of valid blocks referenced by each head, not previously referenced in the node's chain.

Since the small constant $X$ is independent of $N$, each block is of size $\mathcal{O}(1)$, and broadcast to all other $\mathcal{O}(N)$ nodes, ie.\ a communication cost of $\mathcal{O}(N)$. Not all valid blocks need to be included in a node's chain, and a block can be skipped if it is indirectly referenced by a head block included. The interpretation of the block DAG reduces to the basic \sysname: blocks referenced by heads are included indirectly, and interpreted in causal order to reach consensus.

Including $X$ random heads with each block, and a link to the previous block, forms a sparse random graph. With very high probability a block will be included, directly or indirectly, into other nodes chains within $\mathcal{O}(\log N)$ blocks. Since at least three are needed to reach consensus this leads to a latency of $\mathcal{O}(\log N)$.



\section{Related Work}


`Blockchain' designs rely on Nakamoto consensus~\cite{nakamoto2008bitcoin}, originating with Bitcoin, in which a proof-of-work puzzle determines a proposal for a block, and the `most work' rule disambiguate between alternate chains. 
Alternative proposals consider instead mechanisms based on proof-of-stake~\cite{dai1998b}: in such systems, such as Ouroboros~\cite{kiayias2017ouroboros}, nodes are associated with some `stake' value, which is unforgeable and transferable, and decisions on consensus are based on the amount of stake nodes possess. 
All those provide only probabilistic finality guarantees.

Fischer, Lynch and Paterson~\cite{fischer1985impossibility} prove in 1985 that deterministic consensus is impossible under full network asynchrony and in the presence of even a single byzantine fault. Dwork et al~\cite{dwork1988consensus} introduce the partial synchrony model, and present a set of algorithms that achieve byzantine consensus within it. Castro and Liskov~\cite{castro1999practical} introduce Practical Byzantine Fault Tolerance (PBFT) which is deterministic and require partial synchrony for liveness, but is safe under asynchrony. The most mature implementation is SmartBFT~\cite{bessani2014state}, which advertises up to 50K transactions per second (tps). A mixture of those is used today by the tendermint~\cite{kwon2014tendermint,buchman2018latest} smart contract platform. They report a throughput of 10K tps, with consensus reached within 1sec\footnote{\url{https://github.com/tendermint/tendermint/wiki/Introduction}}. Cachin et al.~\cite{cachin2001secure} took the alternative approach and present a probabilistic byzantine consensus scheme based on secure shared randomness.

The worse-case communication cost of the standard PBFT, under repeated view-change, is $\mathcal{O}(N^4)$ where $N$ is the number of nodes in the quorum. This stems from the $\mathcal{O}(N^2)$ size of the new-view messages (containing $2f+1$ prepares from each of the $2f+1$ nodes sending a new-change message, where $2f+1 \sim \mathcal{O}(N)$); the message is sent to all $\mathcal{O}(N)$ nodes; and up to $f \sim \mathcal{O}(N)$ consecutive view changes may be needed before a view with an honest leader is reached. The Tendermint consensus~\cite{buchman2018latest} has a worse case cost of $\mathcal{O}(N^3)$, with cost $\mathcal{O}(N^2)$ per round, and potentially $\mathcal{O}(N)$ rounds with corrupt leaders until a decision is reached.

Abraham et al.~\cite{abraham2018hot}, introduce a linear view change (LVC) tweak to PBFT, to reduce the overall cost to $\mathcal{O}(N^3)$, and further use threshold signatures to reach $\mathcal{O}(N^2)$ including view change. Byzcoin~\cite{kogias2016enhancing}, also employs collective signing to achieve $\mathcal{O}(N)$ only for the steady-state protocol, excluding view change which remains $\mathcal{O}(N^2)$ (each new-view includes $\mathcal{O}(1)$ collectively signed commits, is broadcast to $\mathcal{O}(N)$ nodes, and $\mathcal{O}(N)$ need to be executed to reach a good leader).

Concurrently to our work LinBFT~\cite{yang2018linbft} presents a variant of PBFT with $\mathcal{O}(N)$ communication cost. They achieve this by combining the Linear View Change, from Hot Stuff~\cite{abraham2018hot}, a central point of distribution using Cosi~\cite{kogias2016enhancing} leading to a latency of $\mathcal{O}(\log N)$, and a verifiable random function to determine the leader at every turn (the performance analysis is in terms of asymptotic costs and no specific throughput figures are reported). Byzcoin~\cite{kogias2016enhancing} achieves $\mathcal{O}(N)$ communication complexity for the happy path, but not the view change. Neither of those protocols inherit the beneficial properties of separating the broadcast of block DAG from its interpretation as consensus. 

A few systems interpret DAGs to get finality or consensus. Casper the Friendly Finality Gadget~\cite{buterin2017casper}, provides conditions under which miners' votes lead to a finalized set of blocks. This system influenced Wendy~\cite{abraham2018hot}, which combines a block proposer with a finality layer based on an efficient PBFT variant. The Hashgraph protocol~\cite{baird2016hashgraph} is closest to \sysname in that it also uses a DAG describing transactions sent and received, and a subsequent interpretation step of `virtual voting' to reach consensus without further communication---and identifies the saving in terms of communication complexity that result. \sysname embeds a different interpretation logic, that clearly maps to a simplified PBFT protocol, and as a result does away with the need for electing ``Famous witnesses'' to facilitate consensus, shared randomness generation, and does not require novel consensus correctness arguments. However, the similarity of the approach highlights that different distributed algorithms can benefit from the paradigm of separating the block DAG creation from its interpretation as consensus.

\section{Conclusions}

\sysname illustrates we can interpret a block DAG, constructed based on very weak validity conditions, as instances of a terminating reliable broadcast, and reach consensus on blocks, with safety and finality. Our basic protocol achieves this with worst case communication complexity $\mathcal{O}(N^2)$, and low concrete costs achieving performance of over 400K tps over a wide area network.

The approach of creating block DAGs and interpreting them according to a distributed algorithm, such as consensus, presents great benefits. Network operations, are independent from the state machines of the distributed algorithm, or even higher level protocols such as a proof-of-stake system. Since interpretation is performed by each node independently, there is no need to carry expensive evidence in messages across nodes, leading to efficiency and simplicity. Dispensing of the need of nodes to even know who the other nodes in the consensus are, or the distribution of stake, when creating the block DAG is of particular importance for dynamically re-configurable systems.

We believe that a similar approach may be applicable to interpret block DAGs as crash fail consensus protocols (such as Paxos~\cite{lamport2001paxos} or raft~\cite{ongaro2014search}) or reliable broadcast algorithms without finality. Other byzantine consensus mechanisms, that are probabilistic in nature, such as~\cite{cachin2001secure} may also be adapted by implementing them using collective verifiable randomness. 

\begin{names}
\vspace{3mm}
\noindent {\bf Acknowledgements.} We thank Shehar Bano (UCL) for useful comments on early manuscripts, and Alberto Sonnino (UCL), Mustafa Al-Bassam (UCL), Eleftherios Kokoris Kogias (EPFL), Ben Laurie (Deepmind), Miguel Castro (MSR) and Klaus Kursawe for conversations on the design and its philosophy. Tav Siva and Jeremy Letang ({\tt chainspace.io}) provided experimental validation of the \sysname design, which was useful to understand its implementation implications and performance. George Danezis is supported by the EU H2020 project NEXTLEAP (GA 688722) and EU H2020 DECODE (GA 732546) as well as {\tt chainspace.io}.
\end{names}

{\footnotesize \bibliographystyle{alpha}
\bibliography{sample}}

\appendix

\section{Proofs of Theorems}

\vspace{3mm}
\noindent {\bf Proof of Theorem \ref{dag:avail1}}
\begin{proof}
Since only valid blocks may be included as references into a block emitted, this implies that all their references have been received and stored. The argument applies inductively to all previous blocks, that must also be valid, up to the genesis blocks that contain no past references. \qed 
\end{proof}

\vspace{3mm}
\noindent {\bf Proof of Theorem \ref{dag:avail2}}
\begin{proof}
Since the honest node must have stored the full block DAG referenced by any block it emits, other honest nodes may request any blocks they are missing. Since the emitting node is honest and available they will eventually receive the missing blocks. Note this is irrespective of whether the blocks were created by honest or byzantine nodes. \qed
\end{proof}

\vspace{3mm}
\noindent{\bf Proof of Theorems \ref{trb:safe} \& \ref{trb:live}.} 
\begin{proof}
The proofs of safely and liveness follow closely the arguments for PBFT, as best detailed in~\cite{castro1999correctness}.

The first pre-prepare in view 0, for an instance of the protocol $(n,k)$  has to be signed by node $n$. The need for $2f$ other nodes to broadcast a \emph{prepare} for any pre-preprepared block in a view ensures that only a single block in a view, if any, may ever result in a commit message by any honest node (in effect this is Bracha reliable broadcast~\cite{bracha1985asynchronous}). If that occurs within the timeout delay, then it will result in this unique block being delivered for $(n,k)$, after $2f+1$ commits are received by all --- ensuring safety.

The message \emph{view-change} is triggered by nodes upon a logical or physical time-out, when no progress was made in time. It guarantees liveness, and importantly preserves safety. The \emph{view-change} message contains evidence of the latest block committed, if any. The resulting view change, preserves safety by ensuring that if any node could have received $2t+1$ commit messages for a block $B$ in any past view, then the new-view message will also result in a new view with a pre-prepare for the same value $B$. On the other hand, if no node may have delivered a block in previous views, then the new view will \emph{pre-prepare} `nil'.

If fewer than $f+1$ commit messages were emitted in a previous view, two competing new-view messages may appear in the next view, one for $B$ and one for $nil$. In this case the protocol maintains safety, and is equivalent to a corrupt leader sending two contradictory pre-prepare messages: such a conflict may result in a new view where it will be resolved. Thus the protocol ensures safety and liveness without the need for a leader driving the new-view process. \qed
\end{proof}

\vspace{3mm}
\noindent {\bf Proof of Theorem \ref{int:safe1}}
\begin{proof}
As per Theorem 1, relating to validity, the two nodes will have all the blocks referenced by the block---forming the full DAG from the genesis blocks to this block. Through the collision resistance properties of secure hash functions these sub-graphs of the DAG will be identical in their content. Since the interpretations step is driven entirely by the order blocks are included in other blocks in this sub-graph of the DAG, and is otherwise deterministic, both honest nodes will execute identical steps and reach the same states for the state machines and messages associated with all valid blocks of the DAG. \qed
\end{proof}

\vspace{3mm}
\noindent {\bf Proof of Theorem \ref{int:safe2}}
\begin{proof}
This theorem requires arguing that interpreting the block DAG is equivalent to nodes exchanging messages directly through the terminating reliable broadcast, and therefore the interpretation inherits its safety properties. Consider, without loss of generality an honest node interpreting the state machine relating to a decision for a specific position in a specific block. This honest node will interpret other \emph{honest} node's blocks as messages; and those messages will be exactly the same as if the other nodes sent the terminating reliable broadcast messages directly (see Theorem \ref{int:safe1}). On the other hand, lets assume that the adversary has the ability to inject arbitrary messages into the honest node's interpretation for all blocks that created by \emph{byzantine} nodes (as long as they are authenticated as being sent by one of the byzantine nodes). The result of this strictly more powerful adversary, would be for the state machine interpreted by the honest node to be driven by correct messages from $2f+1$ honest nodes, and arbitrary messages from the $f$ byzantine nodes. 
Therefore the state machine within the blocks of honest user $n$ will follow the same states as if the terminating reliable broadcast was executed explicitly; and the state machines across blocks of two different honest users will therefore deliver the same block for this position (due to the safety properties of the terminating reliable broadcast) as long as there are at most $f$ byzantine nodes. \qed
\end{proof}

\section{View change illustration}

\begin{figure}[t]
\centering
\begin{tikzpicture}[node distance=1.3cm,>=stealth',bend angle=45,thick,scale=0.45, every node/.style={scale=0.8},auto]

\tikzstyle{transition}=[rectangle,thick,draw=black!75,
  			  fill=black!20,minimum size=9mm, text width=1cm]

\begin{scope}
\input{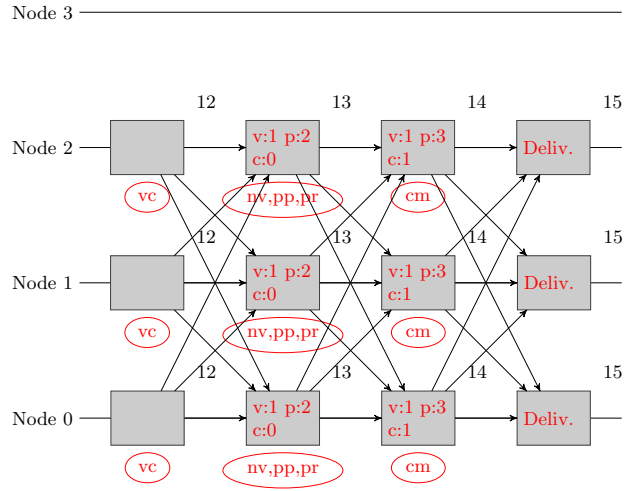}
\end{scope}

\end{tikzpicture}
  \caption{An illustration of a block DAG (black and gray), and its interpretation as a state machine per block for position $(n=3, k=2)$, when node 3 has failed and no block for this position has been broadcast. Each state machine for $(n=3, k=2)$ includes a view number ($v$), and a count for $\textit{prepare}$ ($p$) and $\textit{commit}$ ($c$) messages received by the block, and is illustrated within each block in red (along with a $\textit{Deliv.}$ delivery decision being reached). The $\textit{out}$ buffer for each block for this decision, is summarized in a red circle below the corresponding block: pp for $\textit{pre-prepare}$, pr for $\textit{prepare}$ and cm for $\textit{commit}$, vc for $\textit{view change}$ and nv for $\textit{new view}$. None of the interpreted information (in red) is materialized as network communication.}   \label{fig2}
\end{figure}

Figure~\ref{fig2} illustrates the interpretation of the block DAG in the case a block is missing, and has to be decided as `nil' by honest nodes. In the example node 3 is unavailable, and nodes are called to decide on block $(n=3, k=2)$. They set their timers for view zero at their respective blocks $(n, 2)$, which timeout at their respective blocks $(n, 12)$ where the illustrated trace begins. The honest nodes 0 to 2, then initiate a $\textit{view-change}$ each that they broadcast. Upon those $\textit{view-change}$ messages reaching all nodes (in block round 13) they initiate a $\textit{new-view}$ for $v=1$, that is also interpreted as a $\textit{pre-prepare}$ and $\textit{prepare}$ for the value `nil' for block position $(3,2)$.

The protocol then proceeds as in the steady case, with all nodes sharing the $\textit{prepare}$ messages (received at round 14), followed by the $\textit{commit}$ messages (received at round 15) for the decision $(3,2) \rightarrow $ `nil' to be agreed upon.

\begin{longdoc}

\section{Latency of consensus in blocks}

\begin{figure}[t]
\centering

\includegraphics[width=\linewidth]{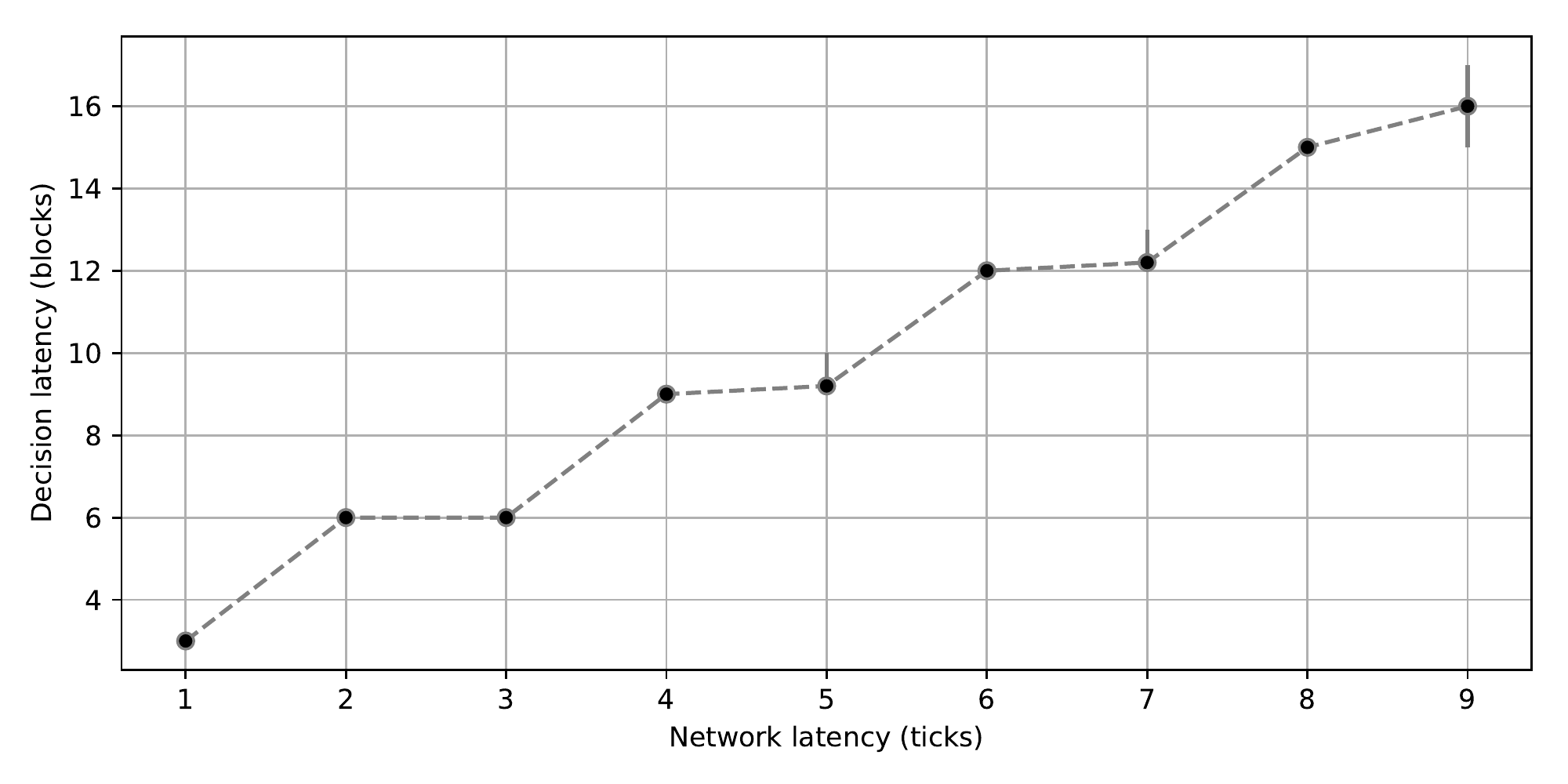}

  \caption{Relationship between network delay and the number of blocks after which a consensus decision is reached. (Block interval is set to $2.0$).}   \label{fig3}
\end{figure}

Figure~\ref{fig3} graphs the relation between latency, and the delay in terms of number of blocks, until a decision is reached for a block. We assume that the the block round interval is $2.0$ ticks, and the network latency is the sum of $l$ and a random sample from an exponential distribution with mean $l/10$. Each data point is the average of 10 runs, and the error bars represent the minimum and maximum value in terms of rounds until consensus is reached.

We observe that when the latency is smaller than the block interval, (such as latency $1.0$ as compared to the set block interval $2.0$) the latency of the consensus is simply 3 rounds of blocks. When the latency increases, consensus is reached in a higher number of rounds. However, the relationship is linear.
\end{longdoc}

\end{document}